\documentclass[]{ccs}
\usepackage{amsmath}
\usepackage{amssymb}
\usepackage{siunitx}    
\usepackage{pdflscape}  
\usepackage{rotating}   
\usepackage{gensymb}    
\usepackage[misc]{ifsym} 
\usepackage{orcidlink}  
\usepackage[formats]{listings}   
\usepackage{alltt}
\usepackage{booktabs}
\usepackage{colortbl}   
\usepackage{tabularx}   
\usepackage{lineno}
\usepackage{longtable}  
\usepackage{subcaption}
\usepackage{multirow}
\usepackage{subcaption}
\usepackage{caption} 
\usepackage{csquotes}   
\usepackage{nameref}
\usepackage{cleveref}
\usepackage[skip=5pt]{caption}
\usepackage{pdfpages}
\lstdefineformat{R}{~=\( \sim \)}
\DeclareSIUnit\Molar{M}

\usepackage[			
	backend=biber,      
    style=apa,   
	natbib=true,		
	hyperref=true,	    
	alldates=year,      
    uniquename=false,   
    maxbibnames=99,     
]{biblatex}
\DeclareFieldFormat{doi}{\mkbibacro{DOI}\addcolon\space\url{https://doi.org/#1}}

\AtEveryBibitem{
    \clearfield{urlyear}
    \clearfield{urlmonth}
    \clearlist{language}
}

\newcommand{\FIG}[1]{\autoref{fig:#1}}
\newcommand{\TABLE}[1]{\autoref{tab:#1}}
\newcommand{\EQ}[1]{\autoref{eq:#1}}

\definecolor{ccsBlue}{HTML}{0066B3}
\definecolor{ccsRed}{HTML}{ed1d30}

\logo{perception}
\logo{LogoCCS}
\logo{tud_logo}

\title{MooneyMaker: A Python package to create ambiguous two-tone images}

\author{Lars C. Reining (corresp.) \orcidlink{0009-0007-7783-5099}}
\affiliation{Technical University of Darmstadt}
\email{lars.reining@stud.tu-darmstadt.de}
\author{Thabo Matthies}
\affiliation{Technical University of Darmstadt}
\author{Luisa Haussner}
\affiliation{Technical University of Darmstadt}
\author{Rabea Turon}
\affiliation{Technical University of Darmstadt}
\author{Thomas S. A. Wallis\orcidlink{0000-0001-7431-4852}}
\affiliation{Technical University of Darmstadt; Center for Mind, Brain and Behavior (CMBB)
Universities of Marburg, Giessen and Darmstadt}

\begin{document}
\maketitle

\abstract[
Mooney images are high-contrast, two-tone visual stimuli, created by thresholding photographic images. 
They allow researchers to separate image content from image understanding, making them valuable for studying visual perception. 
An ideal Mooney image for this purpose achieves a specific balance: it initially appears unrecognizable but becomes fully interpretable to the observer after seeing the original template.
Researchers traditionally created these stimuli manually using subjective criteria, which is labor-intensive and can introduce inconsistencies across studies. Automated generation techniques now offer an alternative to this manual approach. 
Here, we present MooneyMaker, an open-source Python package that automates the generation of ambiguous Mooney images using several complementary approaches. Users can choose between various generation techniques that range from approaches based on image statistics to deep learning models. These models strategically alter edge information to increase initial ambiguity. The package lets users create two-tone images with multiple methods and directly compare the results visually.
In an experiment, we validate MooneyMaker by generating Mooney images using different techniques and assess their recognizability for human observers before and after disambiguating them by presenting the template images. 
Our results reveal that techniques with lower initial recognizability are associated with higher post-template recognition (i.e. a larger disambiguation effect).
To help vision scientists build effective databases of Mooney stimuli, we provide practical guidelines for technique selection.
By standardizing the generation process, MooneyMaker supports more consistent and reproducible visual perception research.
\section{Introduction} \label{intro}
Mooney images are defined as highly degraded two-tone images where each pixel is either black or white, removing all texture and much edge information (see \FIG{example_stimuli} for an example). Because so much information is lost in the generation process, Mooney images are often initially unrecognizable to human observers. This is because it is hard to tell whether an edge present in the Mooney image corresponds to an object boundary or is just the result of a shadow or illumination change in the template image \parencite{mooreRecovery3DVolume1998,hegdeLinkVisualDisambiguation2010}.

This makes Mooney images a useful tool in human vision science. For example, they can be used to study perceptual closure \parencite{mooneyAgeDevelopmentClosure1957,verhallenNewMooneyTest2016,grutznerNeuroelectromagneticCorrelatesPerceptual2010}, holistic processing \parencite{canas-bajoStimulusSpecificIndividualDifferences2020,latinusHolisticProcessingFaces2005}, object recognition \parencite{imamogluChangesFunctionalConnectivity2012,teufelPriorObjectknowledgeSharpens2018,dolanHowBrainLearns1997} or face recognition \parencite{canas-bajoIndividualDifferencesClassification2022,schwiedrzikMooneyFaceStimuli2018,petersonDetectionMooneyFaces2023}. Additionally, Mooney images can also serve practical purposes such as discriminating between humans and machines in CAPTCHA systems \parencite{liMakeUseMooney2024} or user authentication \parencite{castellucciaImplicitVisualMemoryBased2017}.

Mooney images are also used to study the effects of top down processing, because to be able to recognize the content of many Mooney images, observers have to rely on prior knowledge or other forms of top-down processing, and to integrate information over larger spatial regions of the image. This means that often, a Mooney image is only interpretable once the observer has been provided with additional information, e.g., by showing the original template image before the Mooney image \parencite{teufelPriorObjectknowledgeSharpens2018,hegdeLinkVisualDisambiguation2010}.

The ideal Mooney image allows researchers to separate image content from image understanding \parencite{teufelPriorObjectknowledgeSharpens2018}. In other words, the same Mooney image can produce very different percepts depending on whether the observer has prior knowledge about its content or not. This property is interesting for vision science research as it allows to investigate how prior knowledge influences visual perception \parencite{teufelPriorObjectknowledgeSharpens2018,dolanHowBrainLearns1997,hegdeLinkVisualDisambiguation2010}. Thus, an ideal Mooney image can be defined by achieving a specific balance: it should be initially unrecognizable to observers but become fully interpretable after additional information is provided.

While Mooney images have been widely used, they come with the disadvantage of being manually generated. In the study that introduced Mooney images, \textcite{mooneyAgeDevelopmentClosure1957} himself hand-colored image regions black and white to create the images. Nowadays, researchers typically create Mooney images by first smoothing a grayscale image and then applying a global threshold for binarization \parencite{schwiedrzikMooneyFaceStimuli2018,reiningPsychophysicalEvaluationTechniques2024}. Smoothing is typically done by applying a Gaussian filter and is used to remove high-frequency details and noise from the image. During thresholding, the intensity of each pixel (often in the 8-bit range between 0 and 255) is compared to a predefined threshold value. Pixels with intensities above the threshold are set to white (255), while those below are set to black (0). Previously, we have shown that different parameters during smoothing and thresholding can have a significant impact on the interpretability of Mooney images \parencite{reiningPsychophysicalEvaluationTechniques2024}. This is because depending on the choice of these parameters, information relevant for interpretation can either be preserved or lost. This is especially the case if, due to the choice of parameters, edges are distorted which would normally provide important cues like closure for object recognition \parencite{teufelPriorObjectknowledgeSharpens2018}.

As the amount of smoothing and the threshold are typically selected manually and on an image-by-image basis, creating Mooney images is not only time-consuming for large datasets but also introduces variability and potential bias into the generation process. Different researchers might choose different parameters based on their subjective judgment, leading to inconsistencies across studies and laboratories.

To mitigate the impact of these issues, we previously investigated techniques to automatically create Mooney images \parencite{reiningPsychophysicalEvaluationTechniques2024}. For this, we chose predefined smoothing amounts and selected thresholds using modern image thresholding techniques. We found that the amount of smoothing has a larger impact on the interpretability of Mooney images than the choice of thresholding technique.

In this paper, we build upon these findings by developing a set of new Mooney image generation techniques that optimize over both the amount of smoothing and the selected threshold to create Mooney images which are either maximally ambiguous or maximally recognizable. This is achieved by using an optimization procedure that compares the edge maps of the template image and the generated Mooney image using different edge detectors. This decision is motivated by the fact that edges are crucial for object recognition \parencite{biedermanRecognitionbyComponentsTheoryHuman1987} and that, in Mooney images, edge information is distorted as described above \parencite{hegdeLinkVisualDisambiguation2010}.

We then either minimize or maximize the difference between the edge map of the template and the Mooney image by varying the amount of smoothing and the selected threshold. We evaluate these techniques in a psychophysical experiment where human participants had to interpret the generated Mooney images before and after being provided with the template image as a cue. Our results show that the techniques designed to increase the initial ambiguity of Mooney images also lead to a larger increase in interpretability after disambiguation by revealing the template.

All the techniques we developed in this and in our previous study are now part of MooneyMaker, the Python package we introduce here. We provide an easy-to-use implementation of all techniques to generate Mooney images with the goal of enhancing reproducibility and comparability across studies.
\section{The MooneyMaker package} \label{mooneymaker}

\subsection{API overview}
The package offers four core functions. The main function is \texttt{generate\_mooney\_image}, which takes a template image and a selected Mooney image generation technique (described below) as input and returns and saves the generated Mooney image in the specified output directory. Before the Mooney image generation process, the function automatically converts the input image to grayscale and resizes it so that no dimension exceeds 1024 pixels, while maintaining the original aspect ratio. The function also allows users to specify parameters such as the size of the Gaussian smoothing kernel or other technique dependent parameters. Additionally, the function \texttt{convert\_folder\_to\_mooney\_images} allows users to convert all images in a specified folder to Mooney images and store them in an output folder. As described above, users can select the Mooney image generation technique and specify technique dependent parameters.

The function \texttt{plot\_technique\_comparison} takes a template image and generates Mooney images using all available techniques, displaying them side-by-side for comparison. This function is useful for visualizing the differences between the various Mooney image generation methods.

Finally, the package includes the function \texttt{get\_edge\_prediction}. Users can use this function to obtain edge maps of images using the different edge detection algorithms (Canny, TEED, DiffusionEdge) used in the package. This function is particularly useful for understanding how different edge detection methods label the edges in both template and Mooney images.

\subsection{Mooney image generation techniques}

Consider once more that the high-level goal of MooneyMaker is to allow researchers to generate two-tone images that appear ambiguous to humans, but can be readily understood once a template image is shown.
To generate ambiguous two-tone images using global thresholding techniques, we hypothesize that one could find smoothing kernel sizes and thresholds that maximally disrupt the edges of the template image.
In other words, we are seeking a two-tone image whose contours are unlike those of the template image.
(Of course, this also runs the risk of being \textit{too} ambiguous, producing images that are unrecognizable even after the template is shown).

To do this algorithmically, the approach we adopt here is to first detect edges in the template image, then measure the edges in an initial two-tone image, then optimize the thresholding to maximize the difference between the template and the two-tone (i.e. to \textbf{disrupt} the edges).
As comparison conditions to compare using our behavioural experiments, we also include \textbf{similarity} techniques, in which the objective function is reversed and the optimization process finds thresholds to maximize the similarity.

Altogether, MooneyMaker supports nine different techniques for generating Mooney images from template images. Six of these techniques are first introduced here and are optimization-based techniques that either maximize or minimize edge similarity between the Mooney image and the template image. The remaining three techniques are the ones described in our previous work \parencite{reiningPsychophysicalEvaluationTechniques2024}. 

In the following, we describe all nine Mooney image generation techniques implemented in the MooneyMaker package. \TABLE{mooney_techniques} provides an overview of the six novel techniques introduced here.

\subsubsection{Techniques from \textcite{reiningPsychophysicalEvaluationTechniques2024}}
All three techniques described in our previous work \parencite{reiningPsychophysicalEvaluationTechniques2024} use a fixed size of the Gaussian smoothing kernel and differ only in the method used to select the binarization threshold. The size of the smoothing kernel is preselected to be 15x15 pixel but can also be adjusted manually by the user. The size of the smoothing kernel  $k$ relates to the standard deviation $\sigma$ of the Gaussian according to $\sigma = 0.3 \cdot ((k - 1) \cdot 0.5 - 1) + 0.8$. Thus, a kernel size of 15x15 pixels corresponds to a standard deviation of $\sigma = 2.6$ pixels. This preselected standard deviation lies within the range of values used in previous studies \parencite{reiningPsychophysicalEvaluationTechniques2024,keMooneyFaceClassification2017}.\\\\

\noindent\textbf{Mean}\\
This technique uses a fixed size of the Gaussian smoothing kernel and selects the mean pixel intensity of the smoothed image as the binarization threshold.\\

\noindent\textbf{Otsu}\\
This technique also uses a fixed size of the Gaussian smoothing kernel but selects the binarization threshold using Otsu's method \parencite{otsuThresholdSelectionMethod1979}. Otsu's method determines an optimal binarization threshold that minimizes the intra-class variance of the black and white pixel classes.\\

\noindent\textbf{CannyMaxEdge}\\
This technique uses the classical Canny edge detector \parencite{cannyComputationalApproachEdge1986} to find the threshold which maximizes the number of detected edges in the Mooney image. This non-neural technique is described in more detail in \textcite{reiningPsychophysicalEvaluationTechniques2024}. Users can manually adjust the size of the smoothing kernel and the standard deviation of the Canny edge detector. While a small standard deviation of the Canny edge detector will result in fine-grained edge maps, a larger standard deviation will highlight more coarse edges. The default value is 1 pixel. The technique optimizes for the maximum number of edges in the Mooney image. \\

\subsubsection{Novel techniques}
\begin{table}
    \centering
    \begin{tabular}{l l l l}
        \toprule
        \textbf{Technique}      & \textbf{Edge Detection Method} & \textbf{Optimization Objective}              \\
        \midrule
        CannyEdgeSimilarity     & Canny \parencite{cannyComputationalApproachEdge1986}                         & Minimize $L(E_T, E_M)$       \\
        CannyEdgeDisruption     & Canny \parencite{cannyComputationalApproachEdge1986}                           & Maximize $L(E_T, E_M) - R_M$ \\
        TEEDEdgeSimilarity      & TEED \parencite{soriaTinyEfficientModel2023}                        & Minimize $L(E_T, E_M)$       \\
        TEEDEdgeDisruption      & TEED \parencite{soriaTinyEfficientModel2023}                            & Maximize $L(E_T, E_M) - R_M$ \\
        DiffusionEdgeSimilarity & DiffusionEdge \parencite{yeDiffusionEdgeDiffusionProbabilistic2024}                 & Minimize $L(E_T, E_M)$       \\
        DiffusionEdgeDisruption & DiffusionEdge \parencite{yeDiffusionEdgeDiffusionProbabilistic2024}                  & Maximize $L(E_T, E_M) - R_M$ \\
        \bottomrule
    \end{tabular}
    \caption{\textbf{Overview of the six novel Mooney image generation techniques introduced here.} The techniques differ in the edge detection method used to compute the edge maps of the Mooney and template images, the optimization objective (similarity vs. disruption of the template image and Mooney image edge map), and whether regularization is applied. The exact formulation of the loss and regularization can be found in \EQ{loss} and \EQ{regularization}.}
    \label{tab:mooney_techniques}
\end{table}
The following six techniques are first introduced here. All six techniques use optimization to select the size of the Gaussian smoothing kernel and the binarization threshold that either maximizes or minimizes edge similarity between the Mooney image and the template image. The techniques differ only in the edge detection algorithm used to compute the edge maps of the Mooney and template images. An overview of all six techniques is provided in \TABLE{mooney_techniques}.

The optimization is performed using a coarse-to-fine grid search over a predefined range of smoothing kernel sizes and binarization thresholds.
The range from which the size of the smoothing kernel is selected is 11x11 pixels to 39x39 pixels. This preselected range aligns with ranges of values used in previous studies \parencite{reiningPsychophysicalEvaluationTechniques2024,keMooneyFaceClassification2017}.
The range from which the binarization threshold is selected is an intensity between 20 and 190.
Since each image has intensity values in the 8-bit range from 0 to 255, this range excludes very low and very high thresholds that would lead to mostly white or black Mooney images and reduces the size of the search space.

The techniques described in the following optimize for either similarity or disruption of edges between the Mooney image and the template image, using a modified version of the Hausdorff distance \parencite{dubuissonModifiedHausdorffDistance1994}. This objective function measures the similarity between the edge map of a Mooney image and the edge map of the template image and can be described by:
\begin{equation}\label{eq:loss}
    L(E_T, E_M) = \frac{d(E_T, E_M) + d(E_M, E_T)}{2}, \quad \text{where} \quad d(A, B) = \frac{1}{|A|} \sum_{a \in A} \min_{b \in B} \|a - b\|_2
\end{equation}
Here, $E_T$ and $E_M$ are sets of edge pixels in the template and Mooney image respectively, and $\|\cdot\|_2$ is the Euclidean distance between two pixels. The function $d(A, B)$ computes the average distance from each pixel in set A to the closest pixel in set B. The overall loss $L(E_T, E_M)$ is the average of the distances computed in both directions, providing a symmetric measure of dissimilarity between the two edge maps. Since the measure of similarity is symmetric, it not only rewards edges detected in the template image that are preserved in the Mooney image but also penalizes edges that were detected in the Mooney image but not in the template image. The edge maps are computed using a technique specific edge detection algorithm.

Of course, a maximally-disruptive solution would be to make all pixels black or white (i.e. setting very high or low thresholds).
Because we want images that can be disambiguated after seeing the template, this solution would be detrimental.
We therefore subtract a regularization term from the loss in the disruption case. The Mooney image specific regularization term is defined as:
\begin{equation}\label{eq:regularization}
    R_M = \lambda_M \cdot \left|\log\left(\frac{|W_M|}{|B_M|}\right)\right|, \text{where} \quad \lambda_M = \frac{L(E_T, E_M) - \min_{m\in \mathfrak{M}}L(E_T, E_m)}{2}
\end{equation}
Here, $|W_M|$ and $|B_M|$ are the number of white and black pixels in the Mooney image respectively.
In other words, the regularization penalty is higher as the ratio of black to white pixels diverges from one.
In addition, we set the regularization weight $\lambda_M$ as proportional to the distance of the current loss to the minimum loss over all Mooney images for this template image ($\mathfrak{M}$) generated during the grid search.
This ensures that the regularization has a stronger effect when the current loss is high and a weaker effect when the current loss is relatively low.

\begin{figure}
    \centering
    \includegraphics[width=0.8\textwidth]{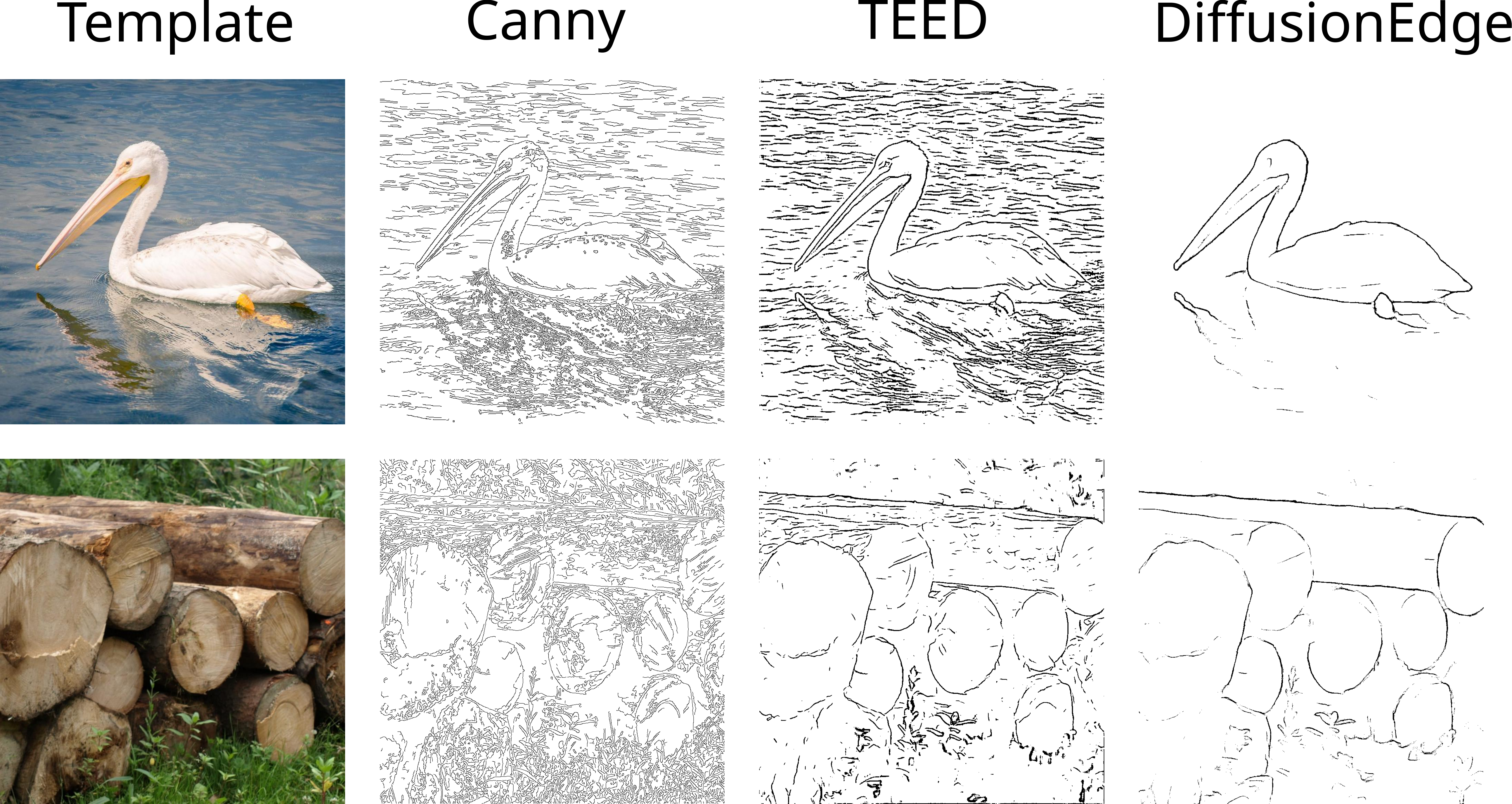}
    \caption{\textbf{Example edge maps of the different edge detection methods.} Shown are edge maps of the same image computed using the three different edge detection methods used in the MooneyMaker package: Canny \parencite{cannyComputationalApproachEdge1986}, TEED \parencite{soriaTinyEfficientModel2023}, and DiffusionEdge \parencite{yeDiffusionEdgeDiffusionProbabilistic2024}. It is easily visible that the different methods capture different kinds of edges in the image. While Canny captures all low-level edges, TEED captures sparser and less texture-like edges, and DiffusionEdge captures mostly object-level edges and contours. The shown images are taken from the THINGS dataset \parencite{hebartTHINGSDatabase18542019}, which is shared under a \href{https://creativecommons.org/licenses/by/4.0/}{CC-BY license}.
    }
    \label{fig:edge_maps}
\end{figure}

The six techniques differ in the edge detection algorithm used to compute the edge maps of the Mooney and template images and whether they optimize for similarity or disruption of edges. Example edge maps computed with the different edge detection methods are shown in \FIG{edge_maps}.\\
\\

\noindent\textbf{CannyEdgeSimilarity}\\
This technique uses the classical Canny edge detector \parencite{cannyComputationalApproachEdge1986} to compute the edge maps of the Mooney and template images. The Canny edge detector is a non-neural technique that detects all low-level edges in the image. The technique optimizes for similarity of edges between the Mooney image and the template image by minimizing the loss $L(E_T, E_M)$.\\

\noindent\textbf{CannyEdgeDisruption}\\
This technique also uses the Canny edge detector \parencite{cannyComputationalApproachEdge1986} to compute the edge maps of the Mooney and template images. However, this technique optimizes for disruption of edges between the Mooney image and the template image by maximizing the loss $L(E_T, E_M) - R_M$.\\

\noindent\textbf{TEEDEdgeSimilarity}\\
This technique uses the Tiny and Efficient Model for the Edge Detection Generalization (TEED) \parencite{soriaTinyEfficientModel2023} to compute the edge maps of the Mooney and template images. TEED is a neural network-based edge detection method that was trained on natural images and is able to detect more semantically meaningful edges. This means, detected edges are sparser and less texture-like compared to the Canny edge detector. The technique optimizes for similarity of edges between the Mooney image and the template image by minimizing the loss $L(E_T, E_M)$.\\

\noindent\textbf{TEEDEdgeDisruption}\\
This technique also uses the TEED \parencite{soriaTinyEfficientModel2023} to compute the edge maps of the Mooney and template images. It optimizes for disruption of edges between the Mooney image and the template image by maximizing the loss $L(E_T, E_M) - R_M$.\\

\noindent\textbf{DiffusionEdgeSimilarity}\\
This technique uses the DiffusionEdge model \parencite{yeDiffusionEdgeDiffusionProbabilistic2024} to compute the edge maps of the Mooney and template images. DiffusionEdge is a state-of-the-art diffusion based model that captures object-level edges and contours in images. The technique optimizes for similarity of edges between the Mooney image and the template image by minimizing the loss $L(E_T, E_M)$.\\

\noindent\textbf{DiffusionEdgeDisruption}\\
This technique also uses the DiffusionEdge model \parencite{yeDiffusionEdgeDiffusionProbabilistic2024} to compute the edge maps of the Mooney and template images. It optimizes for disruption of edges between the Mooney image and the template image by maximizing the loss $L(E_T, E_M) - R_M$.\\
\begin{figure}[t]
    \centering
    \includegraphics[width=\textwidth]{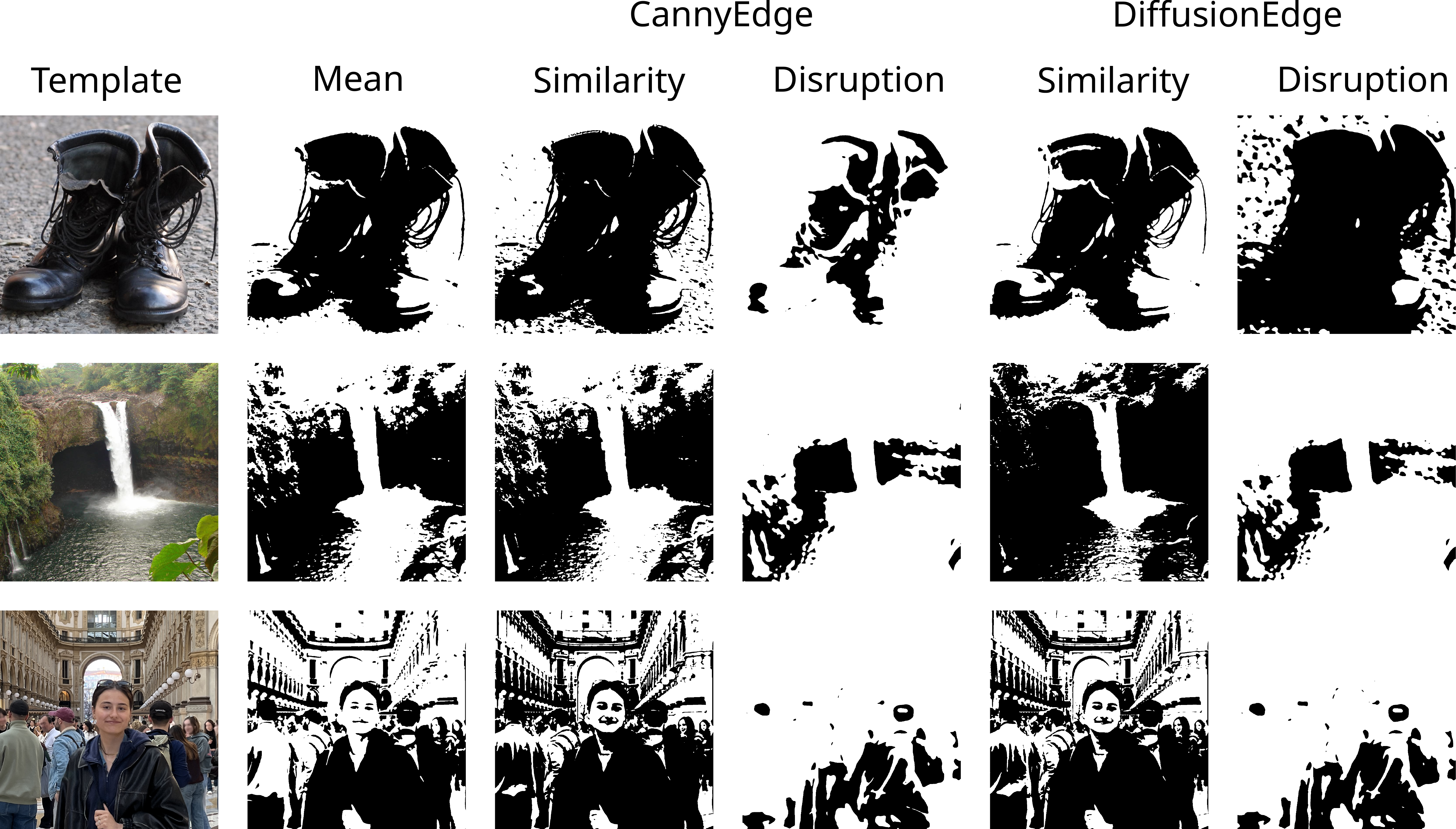}
    \caption{Example stimuli from the three categories (object, natural scene, face) and their corresponding Mooney images generated with the five selected techniques (Mean, CannyEdgeSimilarity, CannyEdgeDisruption, DiffusionEdgeSimilarity, DiffusionEdgeDisruption). The image in the top row was taken from the THINGS dataset \parencite{hebartTHINGSDatabase18542019}. The image in the second row is from the SUN2012 dataset \parencite{xiaoSUNDatabaseLargescale2010}. We took the picture in the third row ourselves, and the depicted person has certified their consent.}
    \label{fig:example_stimuli}
\end{figure}

\subsection{Availability and documentation}
The MooneyMaker Python package is open source (MIT License) and freely available on pypi (\url{https://pypi.org/project/mooney-maker/}) and GitHub (\url{https://doi.org/10.5281/zenodo.18300248}). The repository contains detailed documentation on how to install and use the package, including example notebooks demonstrating its functionality.
\section{Methods} \label{methods}
\subsection{Participants}
Overall, 30 students of TU Darmstadt took part in the experiment. 12 of them were female and 18 male. All of them were naive observers who were not familiar with the purpose of the study. They all had normal or corrected-to-normal vision. All participants certified their informed consent and were rewarded with one participation credit for a course assignment.
All protocols conformed to Standard 8 of the American Psychological Association's Ethical Principles of Psychologists and Code of Conduct (2017) and to the Declaration of Helsinki (with the exception of Article 35 concerning preregistration in a public database).
The experiments reported here were approved by the Technical University of Darmstadt Ethics Commission (Application number EK 77/2022).

\subsection{Stimuli}

\subsubsection{Selection of template images}
As our goal was to test the applicability of the MooneyMaker package to generate Mooney images for different types of images, we selected object, natural scenes and face images as template images. Overall, we used 54 template images, which results in 18 images per category. An example of each category with its corresponding Mooney versions is shown in \FIG{example_stimuli}. Within each category, we also balanced the level of clutter present in each image. As this was not of primary interest for this study, further results on different clutter levels will not be reported here.

Images of objects were selected from the THINGS dataset \parencite{hebartTHINGSDatabase18542019}. An image was categorized as “object” when the object was clearly and prominently depicted, either as the only object in the scene, the largest and most central item, or shown repeatedly to dominate the frame. We preselected images based on this definition, prioritizing objects that are generally recognizable
across individuals.

Natural scene images were selected from the SUN2012 dataset \parencite{xiaoSUNDatabaseLargescale2010} and contained a diverse range of environments, including both natural landscapes (e.g. forests, mountains) and man-made settings (e.g. urban streets, indoor spaces). Unlike object images, which feature a clearly defined item, the selected natural scene images depict broader, more complex environments composed of multiple elements. For example, an image of a café might include chairs, tables, a counter, and background details, making the identification of any single object more difficult and less intuitive. This complexity is characteristic of natural scenes and aligns with their intended role in this study.

Finding an existing dataset suitable for the face category in this study proved challenging, as we not only required high-quality (1000 × 1000 pixels or higher) images of faces, but also a diverse set of backgrounds and lighting conditions to improve generalizability. Additionally, we required images with clearly identifiable emotional expressions, as the experiment involved an emotion recognition task. However, most available datasets with  emotion labels are typically available only under standardized conditions, often with uniform backgrounds and controlled lighting. To ensure greater variability, we therefore created the face stimuli ourselves using an iPhone 14 (iPhone 14 main camera; 26 mm, f/1.5, 9 MP; 3024 × 3024 px; 1:1 aspect ratio). A diverse group of individuals, aged between 18 and 75 years, agreed to participate as models and we obtained permission from all of them for the use of their images. Each person appears
only once in the final dataset, displaying a distinct emotional expression: happy, neutral, angry,
sad or surprised. These expressions are based on the basic emotion framework proposed by \textcite{ekmanAreThereBasic1992}, with a neutral expression added and certain categories omitted. To ensure sufficient variability, the selection included close-up portraits as well as images showing the upper body and larger portions of the background (the main source of clutter in face stimuli). We aimed for diversity in perspective and lighting conditions, incorporating frontal and side views as well as different illumination settings, with the overall goal of representing faces in natural rather than posed environments.
We included faces with different emotional expressions to evaluate human emotion recognition in Mooney images. Thus, we added an additional task: participants had to identify the emotion displayed by each face. However, the dependence on facial emotion expression is not of primary interest here and will not be discussed further.

\subsubsection{Mooney image generation}
We generated Mooney images from each of the 54 templates using five different generation techniques from the MooneyMaker package. As a baseline, we included the Mean technique with a fixed Gaussian smoothing kernel size of 15x15 pixels. In addition, we selected four optimization-based techniques. We choose CannyEdgeSimilarity and DiffusionEdgeSimilarity to represent techniques that optimize for similarity of edges between the Mooney image and the template image. We selected CannyEdgeDisruption and DiffusionEdgeDisruption to represent techniques that optimize for disruption of edges between the Mooney image and the template image. We decided to include both Canny-based and DiffusionEdge-based techniques to compare a classic, non-neural edge detection method with a state-of-the-art diffusion based edge and contour detection model.
All techniques are described in detail in \autoref{mooneymaker}.

Thus, in total, we generated 270 Mooney images (54 template images $\times$ 5 techniques). All stimuli measured 1000 $\times$ 1000 pixels, which, given the experimental setup, corresponded to approximately 12.7 $\times$ 12.7 degrees of visual angle.

\subsection{Procedure and design}
\begin{figure}[t]
    \centering
    \includegraphics[width=0.8\textwidth]{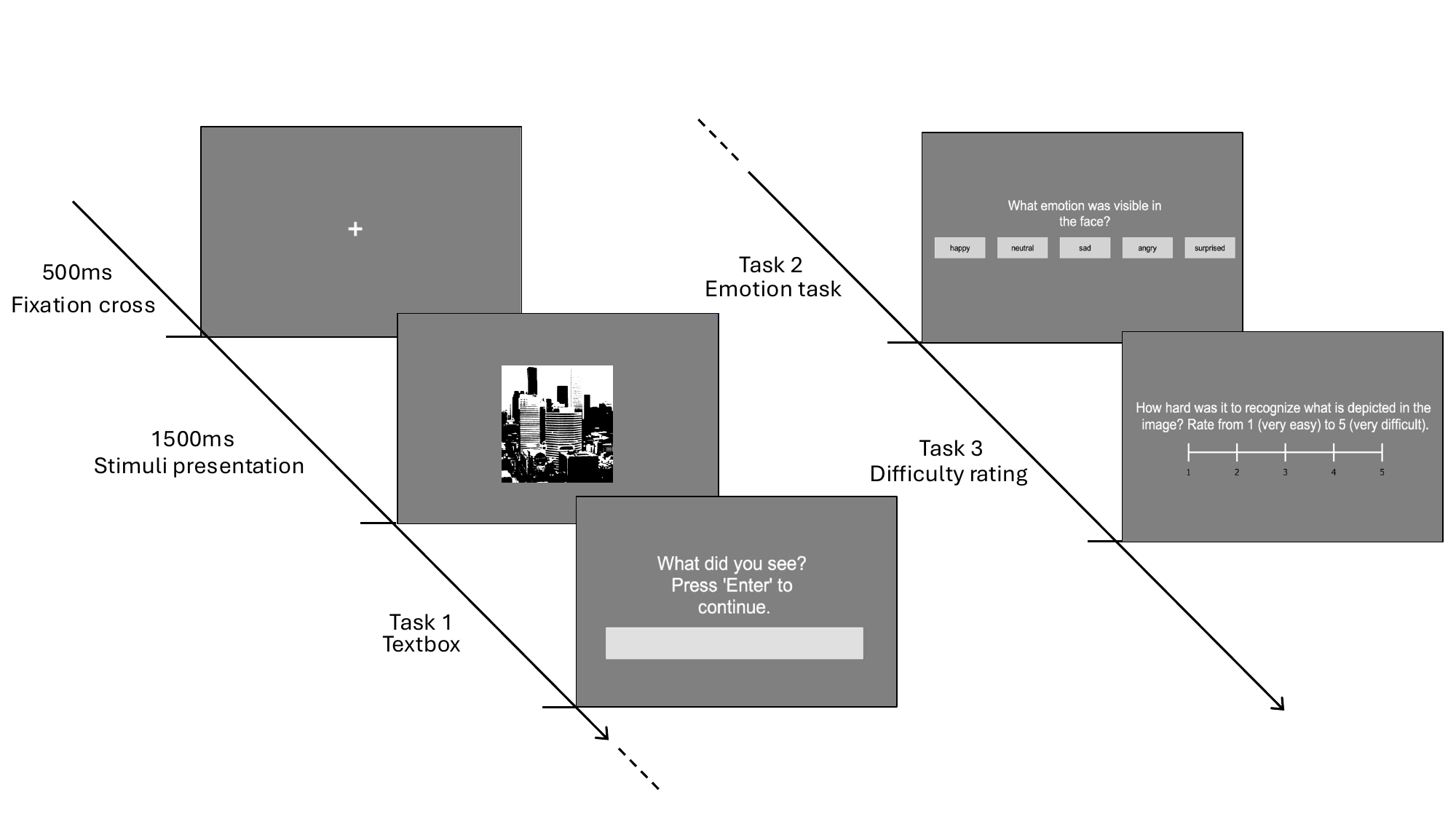}
    \caption{\textbf{Procedure of one trial in one of the two Mooney phases.} Each trial consisted of two to three tasks. First, in the interpretation task, participants were shown a Mooney image and then presented with an open-response text box in which they typed what they saw in the image. If their response indicated that they saw a face, human or person, the emotion recognition task followed, in which they selected one of five emotional expressions (happy, neutral, angry, sad, surprised). If the response did not indicate a face, the emotion recognition task was skipped. Finally, participants were asked to give a subjective rating about the difficulty of interpreting the Mooney image on a scale from 1 (very easy) to 5 (very difficult). Each participant completed two Mooney phases, one before and one after the template presentation phase (not shown here).}
    \label{fig:procedure}
\end{figure}

In this experiment we investigated five different Mooney image generation techniques and their influence on Mooney image disambiguation after template presentation.

The experiment consisted of three phases: two Mooney phases and one template presentation (disambiguation) phase in between. The two Mooney phases are also referred to as pre-template and post-template Mooney phase.
In each Mooney phase, each participant saw 54 Mooney images, one for each template image. The Mooney images were generated using one of the five techniques, randomly assigned to each image for each participant. Thus, no participant saw multiple Mooney images created from the same template image using different techniques. The order of presentation of the Mooney images was randomized for each participant in each Mooney phase. 

Each trial in the Mooney phases followed the procedure shown in \FIG{procedure}. First, a fixation cross was presented for 500 milliseconds to center the gaze of the participants. Then, the Mooney image was presented for 1500 milliseconds. This duration is the same time as used by \textcite{reiningPsychophysicalEvaluationTechniques2024,teufelPriorObjectknowledgeSharpens2018} and should be sufficient to interpret the black and white patches if possible. Immediately afterwards, participants were presented with an open-response text box. They were instructed to type what they saw in the image (interpretation task) by typing one to two words. If their response indicated that they saw a ``face'', ``human'' or ``person'', they were presented with an additional emotion recognition task, in which they selected one of five emotional expressions (happy, neutral, angry, sad, surprised). If the response did not indicate a face, the emotion recognition task was skipped. Finally, participants were asked to give a subjective rating about the difficulty of interpreting the Mooney image on a scale from 1 (very easy) to 5 (very difficult). After this, the next trial of the Mooney phase began. At no point during the experiment did participants receive feedback on their responses. After completing all 54 trials in the first Mooney phase, participants proceeded to the template presentation phase.

The goal of the template presentation phase was to establish a visual connection between the Mooney images and their corresponding template images to allow for disambiguation. To achieve this, the procedure used methods previously used in the study by \textcite{teufelPriorObjectknowledgeSharpens2018}. In each trial of the disambiguation phase, the Mooney image and its original template image gradually faded into and out of each other twice, with each transition lasting three seconds. To ensure that participants maintained their attention throughout the passive viewing, a gray circle appeared at a random location on the image between one and four seconds after the fading sequence ended. The timing of this appearance followed a normal distribution to reduce predictability and prevent participants from anticipating the circle’s appearance. Participants were instructed to click on the circle to proceed to the next trial. 

Following the disambiguation phase, the experiment returned to a second Mooney phase (post-template). The same Mooney images from the pre-template presentation were shown again, but in a newly randomized order. Aside from this reordering, the task remained identical. This design allowed for a direct comparison of perception before and after template presentation, enabling the investigation of how prior exposure to the template influenced interpretation of the Mooney images.

Participants had the opportunity to take short breaks between the three phases of the experiment. The entire experiment lasted approximately 45 minutes per participant, excluding instructions and consent procedures.

\subsection{Equipment}
The code for the experiment and a analysis was implemented in Python (version 3.10.18) using PsychoPy (v2025.1.1). The stimuli were generated using our own MooneyMaker package.

The stimuli were presented on a ROG Swift OLED PG27AQDM monitor, connected to a computer running Ubuntu
22.04. The monitor had a spatial resolution of 3,840 $\times$ 2,160 pixels and a refresh rate of 240 Hz. A chin rest ensured a consistent viewing distance of 70 cm. The study took place in a darkened laboratory at TU Darmstadt.

\subsection{Data analysis}
\subsubsection{Post-Processing of open responses}
To evaluate the open responses from the interpretation task, we first had to post-process and clean the participants answers. Participants (all native German speakers) were instructed to respond in English but were allowed to respond in their German native language if they were not able to recall specific words. We translated all answers to English and corrected spelling mistakes. Furthermore, if multiple words were given by the participant, we tried to extract the most relevant word. This post-processing was done using Google's Gemini 1.5 Flash \parencite{googleGemini2025}. The used prompt is shown in \autoref{app:prompt}. We closely reviewed the post-processed answers to ensure that the meaning of the original response was preserved.

In the next step, we post-processed the ``ground-truth'' labels of the template stimuli. The labeling process
of the template stimuli varied depending on the dataset: images from the THINGS dataset \parencite{hebartTHINGSDatabase18542019} already contained high-quality annotations, while self-created face images were uni-
formly labeled as “face.” For the SUN2012 dataset \parencite{xiaoSUNDatabaseLargescale2010}, some original labels were
slightly simplified (e.g .“access road” was simplified
to “road”). Additionally, in cases where a large number of identical
responses (at least 6) were provided by the participants and these responses clearly and correctly described the image
content but deviated from the label (e.g. "logs" instead of "wood"), we added these as alternative labels for the respective images.

Finally, we evaluated the similarity between the (post-processed) participants' responses and the labels of the template images.
We computed the text embeddings in the pre-trained sentence transformer model "all-MiniLM-L6-v2" \parencite{reimersSentenceBERTSentenceEmbeddings2019}, then calculated the cosine similarity between the response and label  embeddings to obtain a similarity score ranging from -1 to 1, where 1 indicates that the words were identical. However, due to the dense nature of the embedding space, similarity scores were generally non-negative.

This approach allowed us to quantitatively assess how closely each participant's response matched the intended label of the template image. For example, the image in the middle row of \FIG{example_stimuli} shows a waterfall. Pre-template, a participant responded with "plants" which yields a similarity score of 0.35, while post-template, the same participant responded with "river" which yields a similarity score of 0.57. Only the true label "waterfall" would yield a similarity score of exactly 1.0.

\subsubsection{Analysis of interpretation performance and difficulty ratings}
To assess interpretability, we analyzed the similarity scores obtained from the post-processed open responses. Higher similarity scores indicate better interpretability of the Mooney images. We calculated the mean of the similarity scores and ratings for each Mooney image generation technique in both the pre-template and post-template Mooney phases. 

Additionally, we describe the disambiguation effect as the difference in similarity scores/ratings between the post-template and pre-template conditions. 

For all measures we computed 95\% confidence intervals using bootstrapping with 1000 samples. Additionally, we conducted non-parametric Friedman tests to assess overall differences between the five Mooney image generation techniques. If the Friedman test indicated significant differences, we performed post-hoc pairwise comparisons using Wilcoxon signed-rank tests with Bonferroni correction to identify specific differences between techniques.

Correlations between similarity scores and difficulty ratings were assessed using Pearson's correlation coefficient.

\section{Results} \label{results}
\begin{figure}
    \centering
    \begin{subfigure}{0.325\textwidth}
        \phantomsubcaption
        \label{fig:results_similarity}
    \end{subfigure}
    \begin{subfigure}{0.325\textwidth}
        \phantomsubcaption
        \label{fig:results_rating}
    \end{subfigure}
    \begin{subfigure}{0.325\textwidth}
        \phantomsubcaption
        \label{fig:disambiguation_similarity}
    \end{subfigure}
    \begin{subfigure}{0.325\textwidth}
        \phantomsubcaption
        \label{fig:disambiguation_rating}
    \end{subfigure}
    \includegraphics[width=\textwidth]{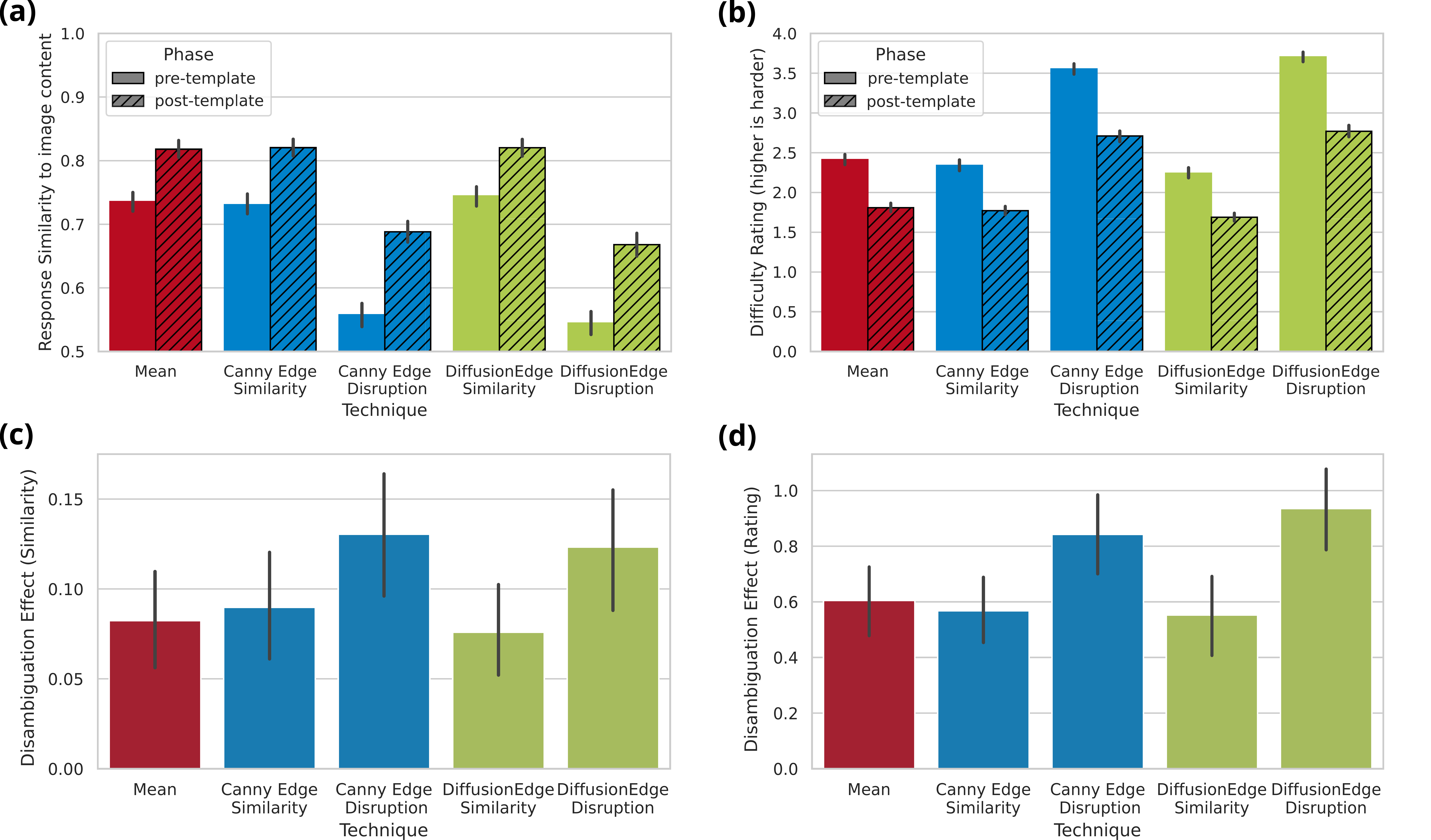}
    \caption{\textbf{Results of the experiment.} For each measure we depict the means with 95\% confidence intervals. \textbf{(a)} Cosine similarity scores between the given responses and the image labels. Higher scores indicate better performance. \textbf{(b)} Difficulty  ratings provided by the participants. Lower ratings indicate an easier task. \textbf{(c)} Disambiguation effect as measured by the difference in similarity scores between the pre-template and post-template conditions. Higher values indicate a larger disambiguation effect. \textbf{(d)} Disambiguation effect as measured by the difference in difficulty ratings between the pre-template and post-template conditions. Higher values indicate a larger disambiguation effect.}
    \label{fig:results_experiment}
\end{figure}

\noindent\textbf{Similar initial interpretability of Mooney images generated with the Mean and Edge Similarity techniques}\\
Across the Mean, CannyEdgeSimilarity and DiffusionEdgeSimilarity techniques, we observed comparable levels of initial interpretability in the pre-template condition. This is shown by the similar similarity scores of the responses (\FIG{results_similarity}) and difficulty ratings (\FIG{results_rating}) provided by participants for Mooney images generated with these techniques. The levels are comparable but not identical. The DiffusionEdgeSimilarity technique shows slightly higher similarity scores and lower difficulty ratings compared to the other two techniques. 

While there is a significant effect of technique on both measures (similarity scores: $\chi^2=58.21$ $p<.001$; difficulty ratings: $\chi^2=76.20$, $p<.001$), post-hoc Wilcoxon tests revealed that the similarity scores and difficulty ratings did not differ significantly between the Mean and edge similarity techniques (all $p_\text{adj}=1.00$).\\

\noindent\textbf{Lower initial interpretability of Mooney images generated with edge disruption techniques}\\
In contrast, Mooney images generated with the edge disruption techniques (CannyEdgeDisruption and DiffusionEdgeDisruption) showed significantly lower initial interpretability in the pre-template condition. This is evident from the lower similarity scores (\FIG{results_similarity}) and higher difficulty ratings (\FIG{results_rating}) compared to the other techniques. Post-hoc Wilcoxon tests confirmed that both measures were significantly different from the same measures for the the Mean and similarity techniques (all $p_\text{adj}<.001$). Furthermore, the DiffusionEdgeDisruption technique led to slightly lower initial interpretability compared to the CannyEdgeDisruption technique, although this difference was not statistically significant ($p_\text{adj}=1.00$).\\

\noindent\textbf{Significant increase of interpretability after template presentation across techniques}\\
Across all techniques, we observed a significant increase in interpretability in the post-template condition compared to the pre-template condition. This is reflected in the positive mean disambiguation effects for both similarity scores (\FIG{disambiguation_similarity}) and difficulty ratings (\FIG{disambiguation_rating}). Additionally, the 95\% confidence intervals for both disambiguation measures did not include zero for any of the techniques, indicating that the increase in interpretability was statistically significant.\\

\noindent\textbf{Similar increase of interpretability after template presentation for Mean and edge similarity techniques}\\
Across the Mean, CannyEdgeSimilarity and DiffusionEdgeSimilarity techniques, we observed comparable increases in interpretability in the post-template condition. This is shown by the similar disambiguation effects for both similarity scores (\FIG{disambiguation_similarity}) and difficulty ratings (\FIG{disambiguation_rating}). While also here there is a significant effect of technique on both disambiguation measures (similarity scores: $\chi^2=13.52$ $p=.008$; difficulty ratings: $\chi^2=20.97$, $p<.001$), post-hoc Wilcoxon tests revealed that the disambiguation effects did not differ significantly between these techniques (all $p_\text{adj}=1.00$).\\

\noindent\textbf{Larger increase of interpretability for edge disruption techniques compared to other techniques}\\
Mooney images generated with the edge disruption techniques (CannyEdgeDisruption and DiffusionEdgeDisruption) showed significantly larger increases in interpretability in the post-template condition. This can be seen by the larger disambiguation effects for both similarity scores (\FIG{disambiguation_similarity}) and difficulty ratings (\FIG{disambiguation_rating}) compared to the other techniques. Nevertheless, the interpretability of Mooney images generated with these techniques after template presentation is still lower than the interpretability of Mooney images generated with the other techniques pre-template. As reported above, there is a significant effect of technique on both disambiguation measures (similarity scores: $\chi^2=13.52$ $p=.008$; difficulty ratings: $\chi^2=20.97$, $p<.001$).\\

\noindent\textbf{Correspondence between objective interpretability and subjective difficulty}\\
The objective measure of interpretability (similarity scores) and the subjective measure (difficulty ratings) showed a consistent pattern across all techniques. Techniques and conditions that resulted in higher similarity scores also tended to have lower difficulty ratings, indicating that participants found these Mooney images easier to interpret (\FIG{results_experiment}). This is supported by a significant negative correlation between the two measures across all pre-template trials (Pearson's $r(1618)=-.58$, $p<.001$) and post-template trials (Pearson's $r(1618)=-.50$, $p<.001$).
\section{Discussion} \label{discussion}
In this analysis, we compared five different techniques to generate Mooney images. 
These techniques and more are now available in the MooneyMaker Python package introduced here. 
We conducted a human experiment to evaluate the interpretability of the created Mooney images before and after disambiguation.

Our results show that Mooney images are generally hard to interpret. Even with techniques that aim to maximize the similarity of edge information between the template and the Mooney image, initial interpretability is not significantly above the level achieved if smoothing is predefined and a mean threshold is used. Neither optimizing for similarity of general edge information (using the Canny edge detector) nor for similarity of object contour information (using the DiffusionEdge contour detector) leads to a significant increase in initial interpretability. Thus, we conclude that simply preserving as much general edge information as possible (as done by the Canny similarity technique) is not sufficient to make Mooney images easily interpretable. This could either be because just too much information is lost in the binarization process of Mooney image generation despite the optimization or because many edges that are preserved might not be relevant for object recognition and could introduce clutter instead which makes interpretation harder \parencite{hegdeLinkVisualDisambiguation2010}.
Additionally, preserving object contour information alone (as done with the DiffusionEdge similarity technique) might not be enough either, as other edge information (e.g., of the environment) could also provide important cues for interpretation.

However, when we use techniques that aim to make Mooney images more ambiguous by disrupting edge information, initial interpretability significantly decreases. This underlines the importance of edge information for recognizing Mooney images \parencite{hegdeLinkVisualDisambiguation2010}. We use two different edge disruption techniques, one based on edges detected with the Canny edge detector \parencite{cannyComputationalApproachEdge1986} and one based on edges detected with the state-of-the-art DiffusionEdge contour detector \parencite{yeDiffusionEdgeDiffusionProbabilistic2024}. While Mooney images generated with the DiffusionEdge disruption technique tend to be slightly harder to interpret initially than those generated with the Canny edge disruption technique, there is no significant difference between the two techniques. This suggests that both edge disruption techniques are effective in reducing the interpretability of Mooney images, likely because they both successfully distort crucial edge information needed for object recognition. This is especially interesting as the DiffusionEdge detector is specifically trained to detect object contours, while the Canny edge detector is a general-purpose edge detector. Thus, both general edge information (this means also edges that are not related to object boundaries) and object contour information seem to be important for recognizing Mooney images.

We were able to achieve a reliable disambiguation effect across all techniques. This means that after showing the template image, the interpretability of Mooney images significantly increased. This confirms previous findings that prior knowledge about the content of a Mooney image can greatly enhance its interpretability \parencite{teufelPriorObjectknowledgeSharpens2018,dolanHowBrainLearns1997,hegdeLinkVisualDisambiguation2010} and that showing the template image is an effective way to provide this prior knowledge.

When comparing the different techniques, we find that the disambiguation effect is significantly larger for Mooney images generated with edge disruption techniques compared to those generated with similarity techniques or with predefined smoothing and mean thresholding. This suggests that Mooney images that are initially more ambiguous benefit more from disambiguation by revealing the template image. This pattern can be explained like this: While edge disruption techniques create Mooney images where crucial visual information is distorted, this information can be recovered when participants are given the template image as a reference. In contrast, Mooney images generated with similarity techniques already contain more recognizable edge information, leaving less room for improvement after disambiguation. Importantly, the smaller disambiguation effect for similarity techniques is not simply due to ceiling effects, since the task remains challenging even after disambiguation.

Finally, we investigated the relationship between objective measures of Mooney image interpretability (i.e. similarity between true label and participant response) and subjective measures (i.e. reported difficulty of interpreting the Mooney image). We find a significant negative correlation between these two measures both before and after disambiguation. This indicates that Mooney images that are objectively easier to interpret are also subjectively perceived as less difficult by participants. This relationship holds true both before and after disambiguation. Notably, the correlation is slightly stronger in the pre-template condition than in the post-template condition. This hints that template presentation might influence perceived difficulty in a way which is different from its influence on actual interpretability.

Disentangling how exactly disambiguation in Mooney images works and which image features are crucial for this process remains an open question for future research. For example, future studies could investigate how different types of prior knowledge (e.g., verbal cues vs. visual templates) influence the disambiguation of Mooney images. Also investigating the influence of image specific effects like image content or statistics (e.g., contrast or spatial frequency) on the interpretability and disambiguation of Mooney images could provide further insights into the underlying mechanisms. With the MooneyMaker package introduced here, researchers now have a tool to systematically create Mooney images with different characteristics, which can be used to address these questions.

Based on our findings, we can provide practical recommendations for researchers choosing between techniques. Techniques that optimize for edge disruption are best suited for creating Mooney images that are initially hard to interpret but show strong disambiguation effects after revealing the template. In contrast, techniques that optimize for edge similarity are more appropriate when the goal is to create Mooney images that are relatively easier to interpret from the start and show moderate disambiguation effects. However, in this case the Mean technique achieves similar results without the overhead of optimization. Therefore, when researchers want higher initial interpretability without strong disambiguation effects, the Mean technique is a reasonable choice.

In conclusion, this study shows that the way Mooney images are created has an influence on their properties in experiments. More specifically, depending on the threshold and smoothing, Mooney images can not only look different but also achieve different interpretability and disambiguation effects. This highlights the importance of reproducible Mooney image generation. For this purpose, we introduce the open-source Python package MooneyMaker which can be used to create Mooney images from digital images. Depending on the use case, researchers should select their technique accordingly.

\section{Acknowledgments}

\subsection{Author contributions}
\noindent\textbf{Lars C. Reining:} Conceptualization, Data curation, Formal analysis, Investigation, Methodology, Project administration, Software, Validation, Visualization, Writing - original draft, and Writing - review \& editing. 

\noindent\textbf{Thabo Matthies:} Methodology, Software, and Writing - review \& editing. 

\noindent\textbf{Luisa Haussner:} Data curation, Investigation, Software, and Writing - review \& editing. 

\noindent\textbf{Rabea Turon:} Conceptualization, Supervision, and Writing - review \& editing. 

\noindent\textbf{Thomas S. A. Wallis:} Conceptualization, Funding acquisition, Methodology, Project administration, Resources, Supervision, and Writing - review \& editing.

\subsection{Funding}

Co-funded by the European Union (ERC, SEGMENT, 101086774). In addition, this work was supported by the Hessian Ministry of Higher Education, Research, Science and the Arts and its LOEWE research priority program ‘WhiteBox’ under grant LOEWE/2/13/ 519/03/06.001(0010)/77 and by the Deutsche Forschungsgemeinschaft (German Research Foundation, DFG) under Germany’s Excellence Strategy (EXC 3066/1 “The Adaptive Mind”, Project No. 533717223). The funders had no role in study design, data collection and analysis, decision to publish, or preparation of the manuscript.
\sloppy
\printbibliography

\appendix
\section{Prompt used to post-process the open responses}\label{app:prompt}
The following prompt was used to post-process and clean the open responses provided by participants in the recognition task using Google's Gemini 1.5 Flash model:

\begin{quotation}
\noindent You will receive a list of human answers from an experiment. 
Most are in English, but some may be in German. Perform the following steps:
\begin{enumerate}
    \item Correct spelling errors. Pay special attention to common German-English keyboard mistakes, especially swapped letters y and z (e.g., "eze" should likely be "eye", "citz" or "citizen" should likely be "city").
    \item Translate German words into English.
    \item Remove leading/trailing whitespace and control characters.
    \item If an answer is verbose or contains multiple words, extract only the core concept.
    \item If an answer contains "or", keep only the first valid option.
    \item If the answer does not contain a meaningful concept, output "NaN".
    \item Make sure the number of outputs equals the number of inputs. Each output should be a single, cleaned concept word.
\end{enumerate}
Example input: ["apple", "nothing", "dont know", "swimsut", "Katze", "besen", "fogel", "stange", "snake (boa)", "satellite image of a storm cloud", "bunnies in a field", "watering can pouring water on plants", "truck or suv from behind"]

\noindent Expected output: ["apple", "NaN", "NaN", "swimsuit", "cat", "broom", "bird", "rod", "snake", "satellite image", "bunnies", "watering can", "truck"]
\end{quotation}

\end{document}